\documentclass[namedreferences]{SolarPhysics}
\usepackage[optionalrh]{spr-sola-addons}
\usepackage{epsfig}
\usepackage{url}

\def\gc{{\gamma}}

\def\bxi{\mbox{\boldmath$\xi$}}

\def\ii{{\rm i}}

\def\br{\mbox{\boldmath$r$}}

\def\unitx{\mbox{\boldmath$\hat{\mathrm{x}}$}}

\def\unitz{\mbox{\boldmath$\hat{\mathrm{z}}$}}
\def\unitr{\mbox{\boldmath$\hat{\mathrm{r}}$}}

\def\ii{{\rm i}}
\def\bk{\mbox{\boldmath$k$}}

\def\bv{\mbox{\boldmath$v$}}
\def\bnabla{\mbox{\boldmath$\nabla$}}
\def\e{\mbox{e}}
\def\bJ{\mbox{\bf{J}}}
\def\bB{\mbox{\bf{B}}}
\def\bF{\mbox{\bf{F}}}
\def\bFL{\mbox{\bf{L}}}

\def\bnabla{\mbox{\boldmath$\nabla$}}

\def\bs{{\cal B}_0}

\newcommand{\eg}{{\it{e.g.}}}
\newcommand{\ie}{{\it{i.e.}}}
\newcommand{\etal}{{\it{et al.}}}
\newcommand{\be}{\begin{equation}}
\newcommand{\ee}{\end{equation}}
\newcommand{\beq}{\begin{eqnarray}}
\newcommand{\eeq}{\end{eqnarray}}

\begin{document}
\begin{article}
\begin{opening}

\title{Helioseismology of Sunspots: 
Confronting Observations with Three-Dimensional MHD Simulations of Wave Propagation}

\author{R.~\surname{Cameron}$^{1}$ \sep
L.~\surname{Gizon}$^{1}$ \sep
T.L.~\surname{Duvall, Jr.}$^{2}$
}

\institute{$^{1}$ Max-Planck-Institut f\"ur Sonnensystemforschung,
                  D-37191 Katlenburg-Lindau, Germany email:\url{cameron@mps.mpg.de}\\
     $^{2}$ Solar Physics Laboratory, NASA Goddard Space Flight Center, Greenbelt, MD 20771, USA}

\runningauthor{R. Cameron \etal}
\runningtitle{Helioseismology of Sunspots}

\date{Received 15 November 2007; accepted }

\begin{abstract}
The propagation of solar waves through the sunspot of AR\,9787 is observed using temporal 
cross-correlations of SOHO/MDI Dopplergrams. We then use three-dimensional MHD numerical simulations to 
compute the propagation of wave packets through self-similar magneto-hydrostatic sunspot models. 
The simulations are set up in such a way as to allow a comparison with observed cross-covariances
(except in the immediate vicinity of the sunspot).
We find that the simulation and the {\it{f}}-mode observations are in good agreement when the model 
sunspot has a peak field strength of 3~kG at the photosphere, less so for lower field strengths. 
Constraining the sunspot model with helioseismology is only possible because the direct effect 
of the magnetic field on the waves has been fully taken into account. Our work shows that the 
full-waveform modeling of sunspots is feasible.
\end{abstract}

\end{opening}

\section{Introduction}
\label{Introduction} 
The subsurface magnetic structure of sunspots is poorly known. The
theoretical picture is that sunspots are either monolithic or
not \cite{Parker79}, or change from being monolithic to ``disconnected'' over
the course of the first few days of their lives \cite{Schuessler05}. 
The observational picture is limited because the subsurface structure is 
inherently difficult to infer. 
The most promising possibility by far is local helioseismology.
Local helioseismology includes several techniques of data analysis,
such as Fourier-Hankel analysis (\eg ~\opencite{Braun95}), time-distance
analysis (\eg ~\opencite{Duvall93}) and helioseismic holography
(\eg ~\opencite{Lindsey97}).
For example the time-distance approach \cite{Duvall93} has been applied to 
determine wave-speed variations and flows associated with sunspots
(\eg ~\opencite{Kosovichev00}; \opencite{Zhao01}; \opencite{Couvidat06}). 
These inversions did not take into account the direct effects of the
magnetic field. We do not mean to give the impression
that such effects have not been investigated, but rather that they are
only beginning to be incorporated into helioseismic inversions. Fourier-Hankel 
decompositions of the acoustic wave field ({\it{p}} modes) near sunspots by \inlinecite{Braun87}
showed that incoming {\it{p}} modes are phase-shifted and ``absorbed'' by sunspots.
This triggered many studies of the effects of the magnetic field on solar waves.
For example, \inlinecite{Spruit91} suggested that sunspot magnetic fields 
are responsible  
for the observed acoustic wave ``absorption'' by partially converting incoming {\it{p}} modes into slow magnetoacoustic waves. This idea was followed up in detail by \inlinecite{Spruit92}, \inlinecite{Cally93}, \inlinecite{Cally94}, 
\inlinecite{Hindman96}, \inlinecite{Cally97}, \inlinecite{Bogdan97}, and \inlinecite{Rosenthal00}. An important result was the 
realization that non-uniform and non-vertical magnetic fields are required to
explain the observations. In particular, 
\inlinecite{Cally00} reported on numerical 
calculations showing that inclined magnetic fields
are able to achieve levels
of absorption compatible with the observations. This aspect of the
problem was followed up by \inlinecite{Cally03} and \inlinecite{Crouch05},
which placed constraints on the strength of the sunspot's magnetic field.
More recently attention has been placed on upward propagating
magnetoacoustic waves and their possible observational signatures 
\cite{Schunker06, Khomenko06, Cally07, Cally07b}.
The use of direct numerical simulations in helioseismology, with or without magnetic fields,
has also undergone rapid development. The aims of these
simulations include validating helioseismic techniques (\eg ~\opencite{Zhao07}) and, more importantly, increasing
our understanding of what the seismic observations reveal about the solar interior.  
Here we restrict our focus to simulations of linearized wave propagation through an inhomogeneous solar atmosphere.
Examples of these linear wave propagation studies, most of which do not include the direct effects of magnetic fields, include   
\inlinecite{Birch01}, \inlinecite{Khomenko06}, \inlinecite{Hanasoge07b}, 
\inlinecite{Hanasoge06}, and \inlinecite{Hanasoge07}.
Some results of these studies clearly have a large bearing on the correct 
interpretation of the helioseismic signatures of sunspots.

In this paper we first design a technique to image the interaction of waves with sunspots, using appropriate averaging of the cross-correlations of the random seismic wave field (SOHO/MDI data). We proceed further by using numerical simulations to do full-waveform forward modeling of the passage of solar waves through a sunspot. This is done with the SliM code \cite{Cameron07}, specifically developed for this purpose.
The ultimate goal is to understand the observed cross-correlations in terms of the properties of our parametric sunspot models.

This paper is organized as follows. The
SOHO/MDI observations of sunspot AR\,9787 are presented in Section~\ref{sec.ar} 
and their helioseismic analysis in Section~\ref{sec.xc}. We describe our numerical simulation code in Section~\ref{sec.code}.
We compare the observations (cross-correlations)  against the vertical velocity on a horizontal cut taken from the simulation at the height of the quiet-Sun photosphere. This, amongst other reasons, means the comparison is only meaningful outside the sunspot.   Except in the immediate vicinity of the sunspot, 
the comparison between the observations and the simulations is very encouraging, as shown in Section~\ref{sec.compare}.
We place a helioseismic constraint on the sunspot magnetic field in Section~\ref{sec.b}.
Our work strongly suggests that we will be able to use observations
and simulations in combination to constrain the subsurface structure of sunspots by taking direct and indirect magnetic effects into account.

\section{SOHO/MDI Observations of Active Region 9787}
\label{sec.ar}

\begin{figure}
\hspace{-0.8cm}
\psfig{file=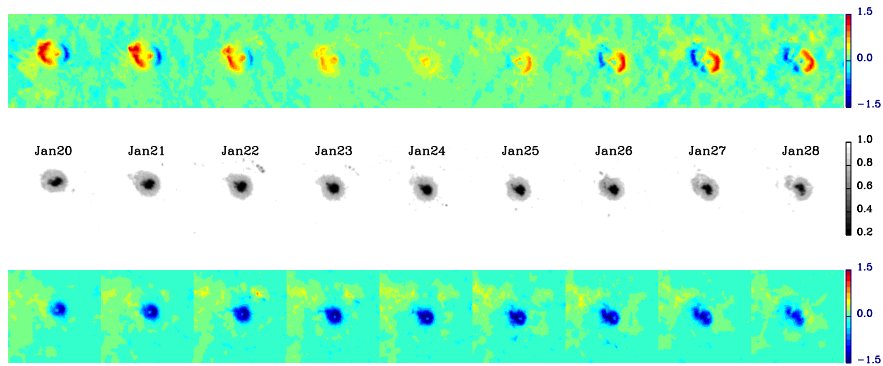,width=12.5cm,clip=,bb=54 360 540 558,angle=0.}
\caption{SOHO/MDI observations of the sunspot of Active Region 9787 during the period 20\,--\,28 January 2002. 
Shown are the daily averages of the line-of-sight Doppler velocity (top row), the intensity (middle row), 
and the line-of-sight magnetic field (bottom row). The color bars are in units of km\,s$^{-1}$, relative intensity, 
and kG, from top to bottom.}
\label{fig.ar9787obs}
\end{figure}

The helioseismic analysis of a sunspot is more or less difficult depending
 on which sunspot is being studied. 
We have been searching the SOHO/MDI database for
the ideal sunspot, \ie a sunspot which is isolated (does not belong to a complex sunspot group), 
has a simple geometry (circular shape), and evolves slowly in time as it moves across
the solar disk. In order to find such a ``theorist's sunspot'', 
we searched the MDI Dynamics Program which combines the advantage 
of a good spatial resolution ($0.12^\circ$ at disk center) with a complete view of the solar disk. 
The full-disk Dopplergrams are available each minute 
during two to three months each year since 1996.
The sunspot we have selected is that of Active Region 9787, continuously observed by MDI during nine days:
20\,--\,28 January 2002. The Dopplergrams were remapped using Postel's azimuthal equidistant projection almost 
-- but not exactly -- centered on the sunspot, using a tracking angular velocity of $-0.1102^\circ$~day$^{-1}$ 
(in the Carrington frame). In addition to the Dopplergrams, we also used the line-of-sight magnetograms 
(each minute) and all the intensity images (one per six hours). The daily averages of these three 
quantities are shown in Figure~\ref{fig.ar9787obs}. Apart from some plage, there is no other active region
in the vicinity of the sunspot during the entire observation sequence. The sunspot of AR\,9787 is large and 
quite stable over the nine days of the observations, even though it starts decaying from 27 January onward. 

\begin{figure}
\psfig{file=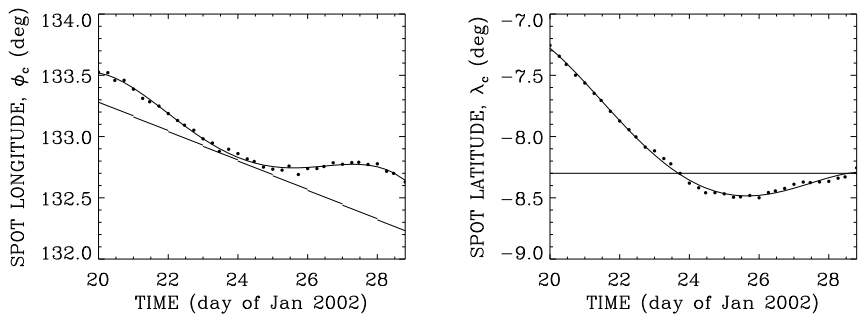,width=12.cm,clip=,angle=0.}
\caption{Coordinates of the center of the sunspot's umbra as a function of time. Left: Carrington longitude 
of the umbra (dots) and Carrington longitude at the center of the Postel projection (line segments). Right: 
Latitude of the umbra (dots) and latitude at the center of the Postel projection (horizontal line). 
}
\label{fig.spotcoordinates}
\end{figure}

\begin{figure}
\centerline{
\psfig{file=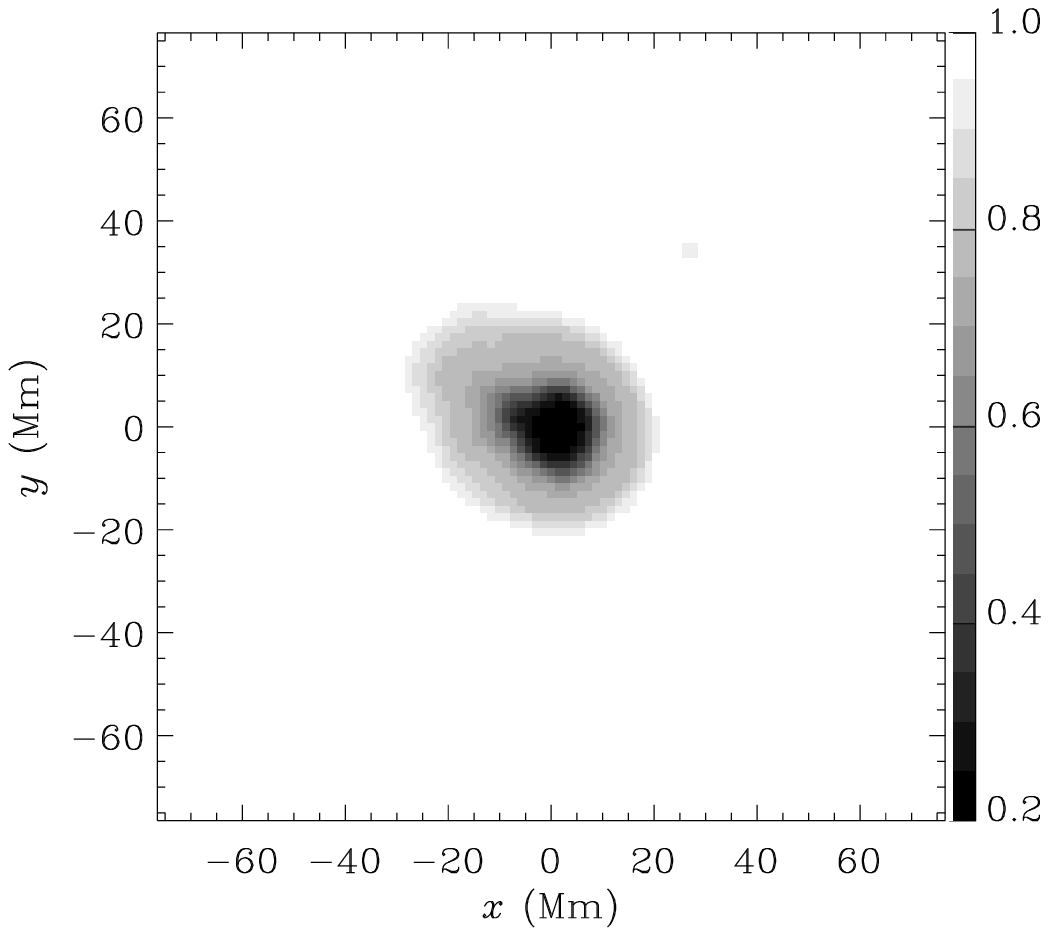,width=6.cm,clip=,angle=0.}}
\caption{SOHO/MDI intensity image of AR\,9787 averaged over nine days, after correcting for the proper motion of the sunspot. 
The intensity is measured relatively to the quiet-Sun value.}
\label{fig.int}
\end{figure}

Using the MDI intensity images, we measured the center of the sunspot's umbra.
As shown in Figure~\ref{fig.spotcoordinates}, 
the sunspot has a significant amount of proper motion. Thus we chose to
split the time series into six-hour subsets and to analyze each subset separately. 
The images belonging to each particular subset were remapped into a new Postel map, 
with the center of the projection corresponding exactly to the center of the sunspot's umbra 
in the middle of the six-hour time interval. 
These 36 six-hour time series of Dopplergrams, centered on the sunspot's position, are the basic data that we used for the 
helioseismic analysis, which we discuss in the next section.  

The average intensity image of the sunspot, corrected for the proper motion of the sunspot, is shown in Figure~\ref{fig.int}. 
From this average image, which is still sharp, we determine the average umbral and penumbral radii to be $9$~Mm and $20$~Mm 
respectively.

\section{Observed {\it{ f}}-mode Cross-Covariance Function}
\label{sec.xc}

We study sunspot AR\,9787 using cross-covariances of the random wavefield, as is done in 
time-distance helioseismology (\opencite{Duvall93}). 
The temporal cross-covariance between two points on 
the solar surface provides information about the Green's function between these 
two points. This interpretation has recently been shown to be correct
in the case of homogeneously distributed random sources in an arbitrarily complex medium
(\opencite{Colin2006}, \opencite{Gouedard2007}).

\begin{figure}
\begin{center}
\psfig{file=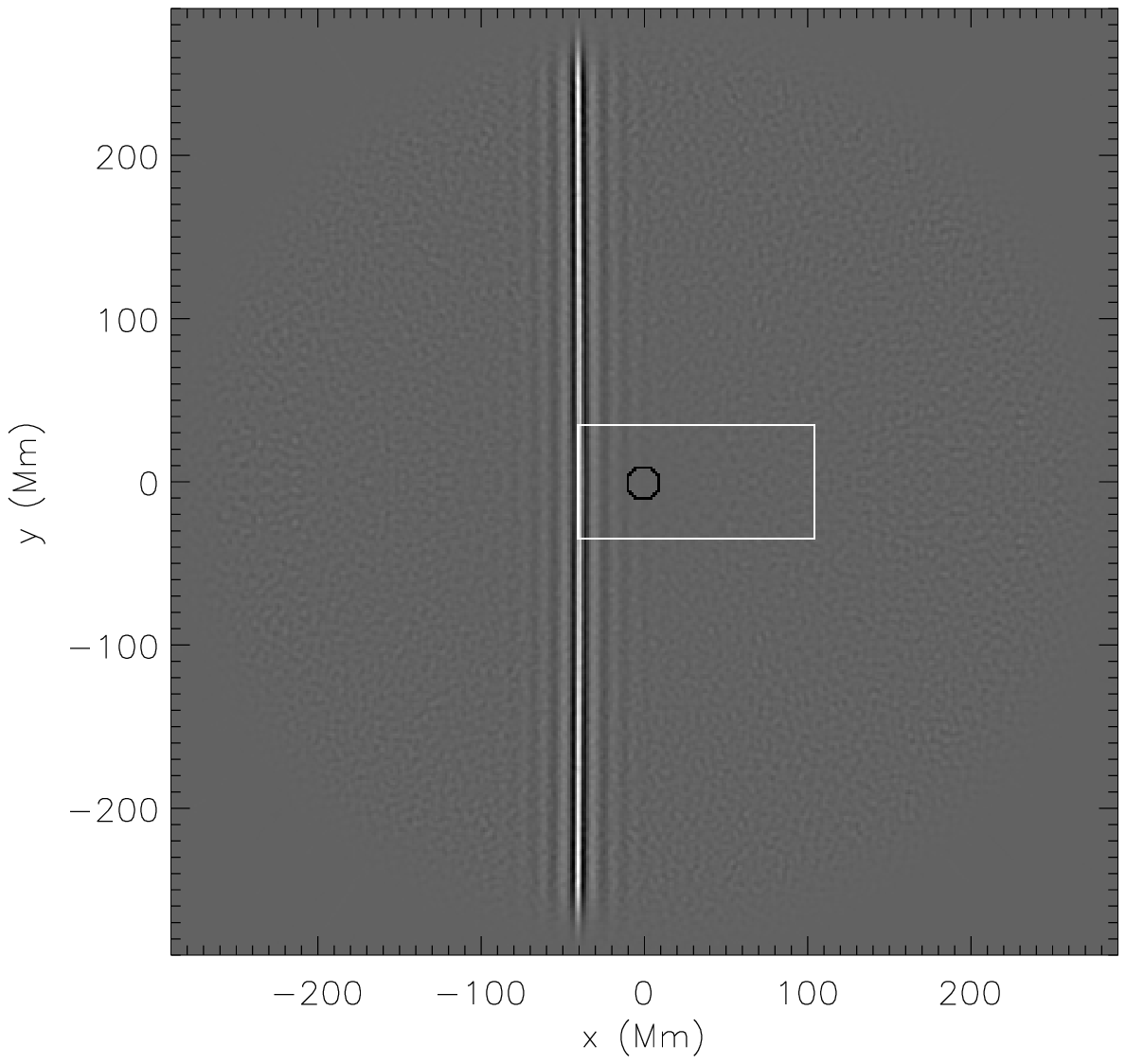,width=12.cm,clip=,angle=0.}
\psfig{file=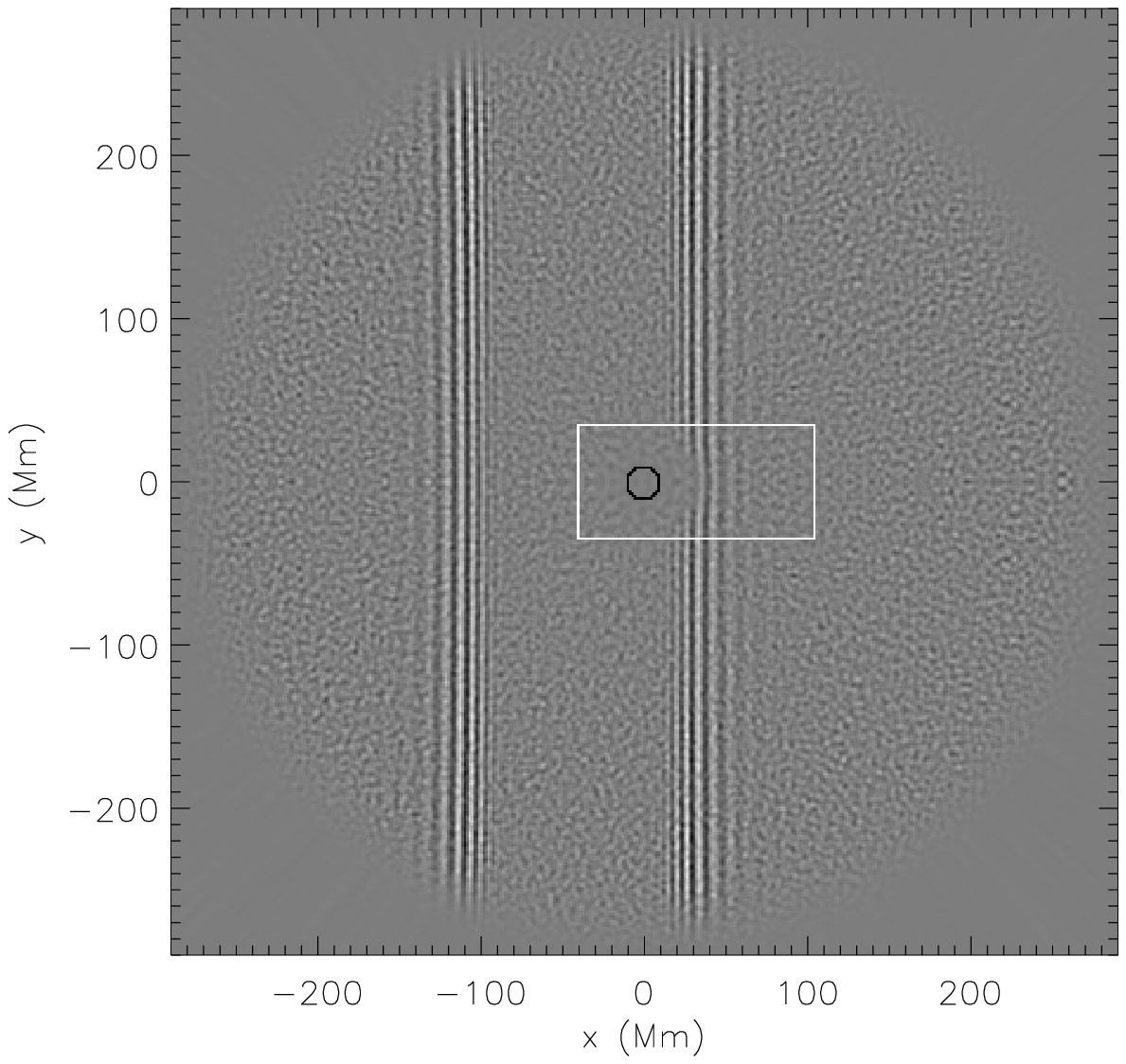,width=12.cm,clip=,angle=0.}
\caption{The upper panel shows the observed {\it{f}}-mode MDI cross-covariance function at zero time lag. 
The sunspot (black circle) is at the center of the Postel map, a distance $\Delta=40$~Mm from the 
meridian $\gc$ (white). The lower panel shows the same cross-covariance at time lag $t=130$~minutes. 
In order to increase the signal-to-noise ratio, the cross-covariance was averaged over all wave 
directions. It is easy to see, by eye, that the waves are affected by their passage through the 
sunspot. The white rectangles show the size of the simulation box.}
\label{fig.obs}
\end{center}  
\end{figure}

In this paper, for the sake of simplicity, we wish to observe the propagation of {\it{f}} modes
through the sunspot. Thus we filter the Dopplergrams in 3D Fourier space to only keep the {\it{f}} modes.
Let us denote by $\phi(\br,t)$ the filtered Doppler velocity, where $\br$ is a position vector and $t$ is time.
Rather than studying the cross-covariance of $\phi$ between two spatial points, we consider the cross-covariance 
between $\phi$ averaged over a great circle $\gc$, denoted by $\overline{\phi}(\gc, t)$, and $\phi$ measured at 
any other point $\bf r$:
\begin{equation}
C(\br,t) = \int_0^T \overline{\phi}(\gc,t') \phi(\br,t+t')  \; {\mathrm{d}} t' ,
\end{equation}
where the effective integration time is $T$ is nine days, or 36 times six hours. Since averaging $\phi$ over $\gc$ is 
equivalent to selecting only the horizontal wavevectors that are perpendicular to $\gc$, the above cross-covariance 
function gives us information about plane wave packets that propagate away from $\gc$. In simple words, the 
cross-covariance at time lag $t$ tells us about the position of a wave packet, a time $t$ after it has left $\gc$. 
In the rest of the  paper we fix the distance between $\gc$ and the center of the sunspot at $\Delta=40$~Mm. 
Since $\Delta \gg \lambda$, where $\lambda\sim5$~Mm is the dominant wavelength, the sunspot is in the far field of $\gc$. 

For any particular choice of orientation of $\gc$, the computed cross-covariance is very noisy. Thus the need 
for some spatial averaging. Let us pick a reference great circle coincident with the meridian at a distance $40$~Mm 
eastward of the center of the sunspot. Because sunspot AR\,9787 is almost rotationally invariant around its center, 
we can compute many equivalent cross-covariance functions corresponding to many different directions of the 
incoming wavepacket, derotate these about the center of the sunspot so that they match the reference cross-covariance, 
and average them to reduce the noise. We have performed this averaging over all the directions of the incoming 
wavevectors, with a fine sampling of 1$^\circ$. As seen in Figure~\ref{fig.obs}, this enables us to reach a very 
good signal-to-noise level.

\section{Three-dimensional MHD Simulation of Wave Propagation through a Sunspot}
\label{sec.code}
\subsection{The Equations}
\label{subsec.eqs}
We use the ideal MHD equations linearized about an arbitrary, inhomogeneous,
magnetized atmosphere. We assume a local Cartesian geometry defined by 
horizontal coordinates $x$ and $y$, and the vertical coordinate $z$ increases
upwards. The level $z=0$ is assumed to correspond to the photosphere, as defined in model S \cite{JCD96}.
 Under the assumptions that gravitational acceleration ($g$) is constant and that there is no background steady flow,
the equation governing the wave-induced displacement 
vector $\bxi$ 
is (\eg ~\opencite{Cameron07})
\begin{equation}
\rho_0 \partial_t^2 \bxi = \bF' ,
\end{equation}
where
\begin{equation}
\bF' =  -\bnabla  P'+ \rho' g \unitz
+ \frac{1}{4\pi} (\bJ'\times \bB_0 + \bJ_0 \times \bB') 
\label{eqn:F} 
\end{equation}
is the linearized force acting on a fluid element (first order in $\bxi$). 
In this paper the subscript $0$ variables represent the steady inhomogeneous background atmosphere and
the primed quantities represent the wave-induced perturbations. In the above equation, the density, pressure, 
magnetic field, and electric current are denoted by the symbols $\rho$, $P$, $\bB$ and $\bJ$ respectively.
The system is closed by the following relations that define $\bF'$ in terms of $\bxi$:
\begin{eqnarray}
\rho'&=&-\bnabla \cdot (\rho_0 \bxi) \label{eq.con} ,\\
P'  &=& c^2_0  ( \rho' + \bxi \cdot \bnabla \rho_0 ) -\bxi \cdot \bnabla P_0 , \label{eqn:energy} \\
\bB'&=&\bnabla \times (\bxi \times \bB_0) , \label{eq.ind}\\
\bJ'&=&\bnabla \times \bB',
\end{eqnarray}
where $c_0$ is the background sound speed.

Whilst we have employed ideal MHD, the waves on the Sun are strongly 
attenuated, presumably as a result of scattering off the time-dependent
granulation. 
Empirically, a solar mode with horizontal wavevector $\bk$ decays as $e^{-\gamma_k t}$ where $1/\gamma_k$ is 
the e-folding lifetime at wavenumber $k=\|\bk\|$.
Introducing ${\bv}'$ as the wave velocity, for each horizontal Fourier mode we write
\begin{eqnarray}
\rho_0 (\partial_t + \gamma_k)\bv'(\bk,z,t) &=& \bF'(\bk,z,t) ,  \label{eq.e1} \\
(\partial_t + \gamma_k) \bxi(\bk,z,t) &=& \bv' (\bk,z,t) .
\label{eq.e2}
\end{eqnarray}
In the above equations, any function $f(\bk,z,t)$ refers to the spatial Fourier transform of $f(x,y,z,t)$.
For $\gamma_k>0$ the system is damped. 
The eigenstates, pairs of $(\bxi, \bv')$, are independent of 
$\gamma_k$, although the eigenvalues
are naturally affected. This is the advantage of
this approach. 
The attenuation that we have implemented acts only in the time domain and is suitable for our purpose, 
although in the Sun attenuation is more complicated.
An alternate way of introducing this phenomenological time attenuation would have been to perform the ideal 
calculation and then impose the decay in the time domain {\it{post-facto}} after the 
calculation has ended.

In this paper we focus on the {\it{f}} modes (solar-surface gravity waves), for which the attenuation has been measured to be
\begin{equation}
\gamma_k = \gamma_* \left(k/k_*\right)^{2.2} ,
\end{equation}
where  $\gamma_*/\pi=100$~$\mu$Hz and $k_*=902/R_\odot$ is a reference wavenumber \cite{Duvall98,Gizon02}.

\subsection{The Code}
We use a modified version of the SLiM code which is, apart from the 
modifications discussed below, described in \inlinecite{Cameron07}. The code
has been tested against analytic solutions -- some of these tests are also
described in detail in \inlinecite{Cameron07}. 

The present code includes two absorbing layers at the top and the bottom of the box. 
In the top layer above the temperature minimum we heavily damp the waves and systematically reduce the effect 
of the Lorentz force. This layer only affects waves which have 
escaped through the photosphere.  Likewise the bottom layer damps the waves that propagate downwards.
The purpose of both layers is to minimize the effects of the boundaries
 which would otherwise artificially reflect the waves.

An additional change has been to introduce the mode attenuation described in Section~\ref{subsec.eqs}. 
Since we use a semi-spectral scheme the implementation was straightforward. 

The scheme uses finite differences in the vertical direction, with 558 uniformly
spaced grid points sampled at $\Delta z = 25$~km. 
In the horizontal direction we use a Fourier 
decomposition with 200 modes in the $x$ direction and only 50 in the $y$. The spatial sampling 
is $\Delta x = 0.725$~Mm and $\Delta y = 1.45$~Mm. This
seemingly low resolution is satisfactory because neither the initial wave-packet
nor the sunspot has any significant power at short wavelengths. The size of the
simulation box is $145$~Mm long ($x$-coordinate), $72.5$~Mm across ($y$-coordinate) and $14$~Mm 
in depth ($12.5$~Mm below the photosphere). A typical run, such as the one from Section~\ref{sec.waveform}, 
takes approximately 14 days on a single CPU core.

\subsection{Stabilizing the Quiet-Sun Atmosphere}
\label{sec.qs}
Our aim is to model the propagation of waves through the solar atmosphere
in such a way as to allow a direct comparison with the observations. The
most direct way of proceeding would be to use an existing solar-like atmosphere such
as that of model S \cite{JCD96}. 
This most direct approach is however unavailable when considering
the full evolution of wavepackets using numerical codes because both the solar and 
model S atmospheres are convectively unstable.

Wave propagation through unstable atmospheres is difficult to study 
numerically because any numerical noise in the unstable modes grows
exponentially and eventually dominates the numerical solution. The problem is
then to stabilize the atmosphere whilst leaving it as solar-like as possible,
at least in terms of the nature of the waves propagating through it.

The convective instability is easily understood by reference to Equation~(\ref{eqn:energy}). 
We imagine a small vertical displacement of a blob of plasma,
and assume it is evolving slowly enough that it is in pressure balance with its
surroundings. This implies $P'=0$ and hence 
$c_0^2 \rho'=\bxi \cdot \bnabla P_0 - c_0^2 \bxi \cdot \bnabla \rho_0$. 
For a vertically stratified atmosphere, this becomes 
$\rho'=\xi_z (\partial_z P_0 - c_0^2 \partial_z \rho_0)/c_0^2$. The atmosphere
is convectively unstable if an upward displacement ($\xi_z>0$) corresponds to
a region of lowered density ($\rho'<0$); in such a case the fluid parcel is 
buoyant and accelerates upwards. The atmosphere is convectively stable 
when $\xi_z$ and $\rho'$ have the same sign, requiring 
$(\partial_z P_0 - c_0^2 \partial_z \rho_0)/c_0^2>0$, equivalently 
\begin{eqnarray}
\partial_z P_0 > c_0^2 \partial_z \rho_0. 
\label{eqn:stability}
\end{eqnarray}

We have the freedom to modify any combination of $P_0$, $c_0$, or $\rho_0$ 
in order to satisfy Equation~(\ref{eqn:stability}). It is somewhat natural to regard 
these three quantities as being related through an equation of state or through
the constraint that the atmosphere be hydrostatic, however neither of these
relationships is necessary in terms of the properties of the propagating waves.
We choose $P_0$, $c_0$, and $\rho_0$ so that the wave speed
is solar-like. Changing the sound speed would obviously have a major
impact on the propagation of sound waves, and hence we have chosen to
keep $c_0$ unchanged. Varying $\rho_0$ would have the effect of varying the
distribution of the kinetic energy density of the different modes as a function of
height, which would significantly change the sensitivity of wave packets to 
inhomogeneities; so we have also kept $\rho_0$ fixed. Thus we choose $P_0$ so that
\begin{eqnarray}
\partial_z P_0=\mbox{max}\{0.9 c_0^2 \, \partial_z \rho_0, \; \partial_z P_{\rm u}\} ,
\end{eqnarray}
where  $P_{\rm u}$ was the pressure distribution of the unstable atmosphere.
Notice that $\partial_z P_0$ is negative so that the factor of $0.9$
does indeed mean that Equation~(\ref{eqn:stability}) is satisfied.
The factor of 0.9 is possibly unnecessary, but does little harm: setting this
factor to 1 should create stationary eigenmodes, setting it to $0.9$ introduces
internal gravity modes which propagate very slowly.

\subsection{An Example Oscillation Power Spectrum}
\label{subsec.power}

\begin{figure}    
\centerline{\psfig{file=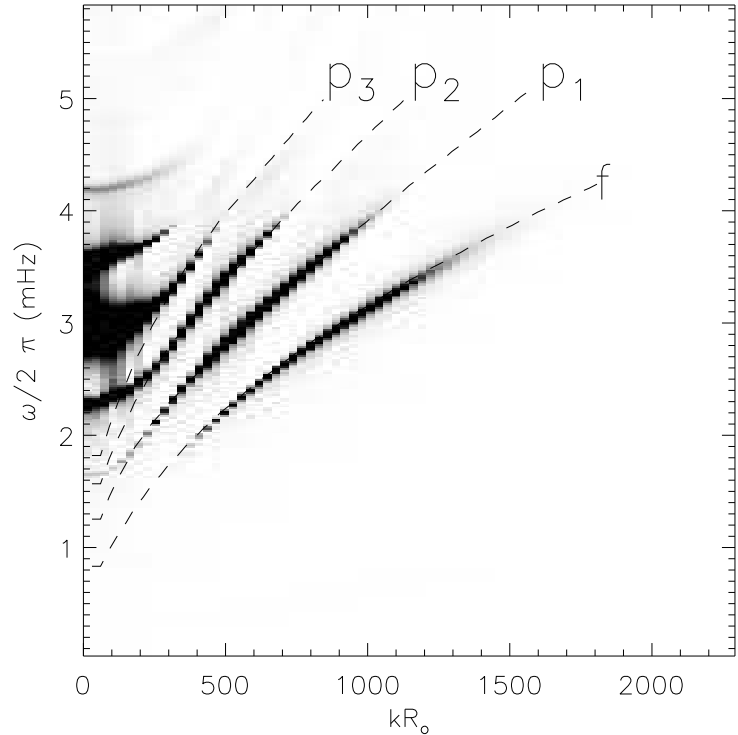,width=17.cm,clip=}}
\caption{Power as a function of wavenumber and frequency for the example simulation described in 
Section~\ref{subsec.power} (no sunspot). The dashed lines correspond to the 
eigenfrequencies of the model S atmosphere.}
\label{fig:k_omega}  
\end{figure}

Since we have modified the atmosphere, we need to check whether it supports oscillations 
that are those of the Sun and model S in the absence of a sunspot. We chose initial conditions 
consisting of a ``line source'' of  the form
\begin{equation}
\xi_z=\e^{-{x^2}/2 s^2} \e^{-{(z-z_0)^2}/2 s^2} \quad {\rm at} \; t=0 ,
\end{equation}
where $s=0.7$~Mm and $z_0=-250$~km is an arbitrary height below the photosphere. All other wave 
perturbations, $\xi_x$, $\xi_y$, and $\bv'$ being zero at $t=0$.
This disturbance encompasses many different modes allowing several ridges of the dispersion diagram to emerge. 
Figure~\ref{fig:k_omega} directly compares these ridges with the eigenfrequencies of model S 
(A.C. Birch, private communication).  The {\it{f}} modes and the acoustic modes p$_1$, p$_2$, and p$_3$ lie 
where they are expected for $k R_\odot>300$. 
It is also to be noted that there is a very low amplitude ridge (not visible on the plot) at frequencies 
below 1~mHz that corresponds to internal gravity modes,  which are not present in the Sun. 
They are an artifact of having made the system convectively stable as explained above, but they 
can be safely ignored.

The simulations and model S eigenfrequencies differ strongly at wavenumbers less than $300/R_\odot$. 
This difference is due to the fact that our computational domain extends to only 12~Mm below the 
solar surface. We have not tested a deeper box because the {\it{f}}, {\it{p}}$_1$, and {\it{p}}$_2$ modes 
look satisfactory in the range of wavelengths we are interested in ($k > 300/R_\odot$).

\subsection{A parametric sunspot model}
\label{sec.sunspot}
In this section we will describe the sunspots that we have embedded in our 
atmosphere. We begin by noting that we embed the sunspot in the atmosphere before 
we stabilize it, in order to get the correct sound speed and density structures.
We use $P_{\rm u}$ to denote the pressure of the atmosphere {\it before} it
has been stabilized as described above.
The sunspot is modeled by an axisymmetric magnetic field $\bB_0 = \bB_0(r,z)$, where $r$ 
is the horizontal radial distance from the sunspot axis.
Within the part of the atmosphere described by model S, the magnetic field is made 
hydrostatic in a standard way, by calculating the Lorentz force 
$\bFL = (\bJ_0 \times \bB_0)/4\pi$ and noting that
the horizontal force balance then requires a horizontal pressure gradient: 
$\partial_r P_{\rm u} = \unitr\cdot\bFL$.
Since $P_{\rm u}$ is unaffected by the spot at large distances, we can integrate from 
infinity towards the center of the spot to find $P_0$. Having thus found $P_u$ we 
can find $\rho_0$ by the constraint of vertical force balance. In this case the 
gravitational force needs to balance both the vertical component of the Lorentz 
force as well as that of the pressure gradient. In principle given $\rho_0$ and $P_0$
the Saha-Boltzmann equations need to be solved to obtain the first adiabatic exponent ($\Gamma_1$)
and the sound speed ($c_0$). However for the purposes of this paper we have assumed
that $\Gamma_1$ is a function of $z$ only and is unaffected by the sunspot. This
assumption will be relaxed in future studies. 

The procedure thus outlined can always be applied, however in some cases the 
results will involve negative pressures or densities. Such solutions are of course
unphysical and indicate that no hydrostatic solution exists for the given 
magnetic configuration and quiet-Sun pressure stratification. The problem 
typically arises in the very upper layers of the box when the magnetic pressure
is large. For example a purely vertical flux tube with a magnetic pressure $P_{\rm m}$
cannot be in hydrostatic balance in an atmosphere with external pressure
$P_{\rm ext}<P_{\rm m}$. In practice the Sun rapidly evolves
to almost force-free field configurations above the solar surface. The role of
the Lorentz force in structuring the atmosphere in the low-$\beta$ region is thus 
artificial as well as being responsible for the lack of equilibrium. It is also 
dynamically of minor importance to the waves since it is above both the acoustic 
cut-off and the layer where the acoustic and Alfv\'en wave speeds become equal. We have
therefore adopted the approach (\AA. Nordlund, in the context of realistic 
photospheric magnetoconvection simulations of active regions, private communication) 
of scaling the Lorentz force
in this region when constructing the background atmosphere. The scaling factor was 
chosen to be $1/[1+ B_0^2/( 8 \pi P_{\rm qs})]$, where $P_{\rm qs}$ is the quiet-Sun 
pressure in the absence of the spot. This scaling factor works, although 
a more conservative scaling will be tested in the future.

Above the model S atmosphere, 
the density and pressure fall rapidly and hydrostatic balance requires an extreme
scaling of the Lorentz force. However since we are not aiming to realistically model 
waves that reach these heights (we only want to damp them) this is acceptable. Instead 
of requiring force balance in this purely artificial region, we have chosen to  put
$P_0(x,y,z)=P_{\rm qs}(z) P_0(x,y,z_0)/P_{\rm qs}(z_0)$ where $z_0$ is the top of the model S
atmosphere and $P_{\rm qs}(z)$ is the pressure stratification of the system in the
absence of the spot (the quiet-Sun value). The density was treated similarly.

In this paper we follow \inlinecite{Schlueter58},   \inlinecite{Deinzer65}, and 
\inlinecite{Schuessler05} amongst others in concentrating on axisymmetric 
self-similar solutions. For this class of models, the vertical component of the magnetic field
 is assumed to satisfy
\begin{equation}
B_{0z}(r,z)=\bs \, Q\left(r\sqrt{H(z)}\right) \,  H(z) ,
\end{equation}
where the functions $Q$ and $H$ satisfy $Q(0)=H(0)=1$ but are 
otherwise arbitrary functions, and $\bs$ is a scalar measure of the vertical-field
strength at the surface. The usual 
practice \cite{Solanki03}, which we adopt in this paper, is to assume $Q$ is a Gaussian,
\begin{equation}
Q(r)=\e^{- (\ln 2) r^2/R_0^2} ,
\end{equation}
where $R_0$ is the half-width at half-maximum.
We chose $R_0=10$~Mm to
correspond to the observed value for sunspot AR\,9787, since $R_0$ is the half-width at half-maximum of the model 
sunspot at the surface.
The value of $R_0$ is  fixed throughout this paper. For the function $H$ we chose 
the exponential function 
\begin{equation} 
H(z)=\e^{z/\alpha} ,
\end{equation}
 with $\alpha=6.25$ Mm. This choice is rather arbitrary and will certainly be varied in the near future. 
In this paper we concentrate on varying $\bs$, the peak magnetic field at the surface. Having thus 
prescribed the formula for $B_{0z}(r,z)$, we determine $B_{0r}(r,z)$ by the requirement that $\bnabla \cdot \bB_0=0$. 
Once we have such a hydrostatic solution, we adjust the pressure everywhere 
to make it convectively stable in the manner described above.

\section{Comparison between Simulations and Observations}
\label{sec.compare}

In this section we want to compare the observed {\it{f}}-mode cross-covariance from AR\,9787 and the simulations.  
The computational domain is $-40$~Mm$<x<105$~Mm, $-36.25$~Mm$<y<36.25$~Mm, and   $-12.5$~Mm$<z<1.5$~Mm. 
The sunspot axis is $x=y=0$. The relationship of the computational box to the observations is shown by 
the white rectangle in Figure~\ref{fig.obs}.

\begin{figure}    
\begin{center}
\psfig{file=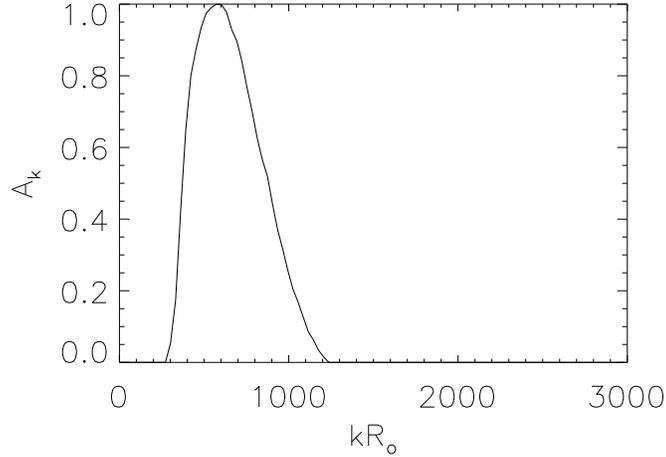,width=10.cm,clip=,angle=0.}
\caption{The initial distribution of {\it{f}}-mode amplitudes used in the simulations, $A_k$, as a function of $k R_{\odot}$.}
\label{fig:obs_spectra}
\end{center}  
\end{figure}

\subsection{The Initial Conditions for the Simulation}
\label{sec.simu}
In this section we discuss the choice of the initial conditions for the simulation. The aim is to allow a direct comparison
between the simulated wave field (Section~\ref{sec.code}) and the observed cross-covariance function (Section~\ref{sec.xc}).   
Since the observed cross-covariance uses the $z$ component of the wave velocity as input data, we choose
to use $v_z'$ from the simulations for the comparison (this is also in accordance with a deeper interpretation of the 
cross-covariance, see \opencite{Campillo03}).  The vertical component of velocity of any {\it{f}}-mode wave packet propagating 
in the $+\unitx$ direction in a horizontally homogeneous atmosphere is of the form
\begin{equation}
v_z'(x,y,z,t)=\mbox{Re} \sum_k A_k \e^{kz} \e^{\ii k (x-x_0) - \ii \omega_k t -\gamma_k t} ,
\label{eq.vz}
\end{equation}
where $A_k$ are complex amplitudes, and $\exp(k z)$ and $\omega_k=\sqrt{g k}$ are the {\it{f}}-mode eigenfunction 
and eigenfrequency at wavenumber $k$. We introduced the reference coordinate $x_0=-\Delta$,
where $\Delta=40$~Mm is defined in the previous section as the distance between $\Gamma$ and the sunspot.
Given the evolution Equations~(\ref{eq.e1}) and~(\ref{eq.e2}), this wave packet is uniquely determined by the initial conditions
\begin{eqnarray}
\bv' &=&  {\rm Re} \sum_k (\ii \unitx + \unitz) \,  A_k \e^{k z+\ii k (x-x_0)} , \\
\bxi &=&  {\rm Re} \sum_k (-\unitx + \ii\unitz) \, \omega_k^{-1}  A_k \e^{k z+\ii k (x-x_0)} .
\end{eqnarray}
Thus our problem is reduced to fixing the amplitudes $A_k$.  In this study, the amplitudes $A_k$ are 
real (all waves are in phase at $x=x_0$ and $t=0$) and are shown in Figure~\ref{fig:obs_spectra}. 
This choice was simply the result of requiring that the simulated $v_z'(x,y,z=0,t)$  and the 
cross-covariance look approximately the same, in the far field and in the absence of the sunspot.  
As will be shown in the next section, this spectrum of wavenumbers is sufficiently accurate 
for the present study.  A more systematic analysis is planned. One important difficulty that arises is the existence 
of background solar noise which is not easy to model and has been ignored in Equation~(\ref{eq.vz}).

\subsection{Magneto-Acoustic Waves}

\begin{figure}    
\begin{center}
\psfig{file=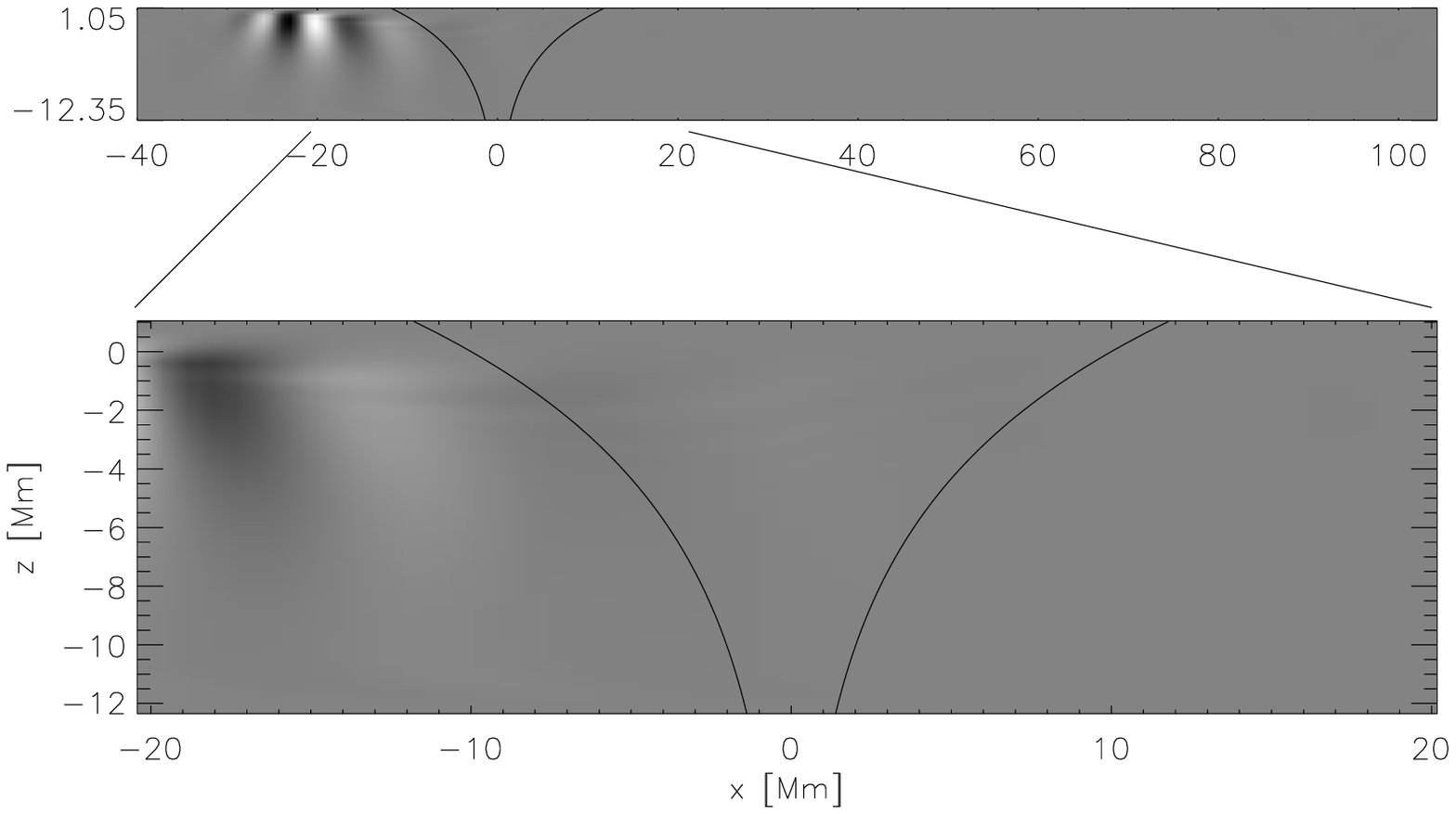,width=12.cm,clip=,angle=0.}
\vspace{-1cm}\psfig{file=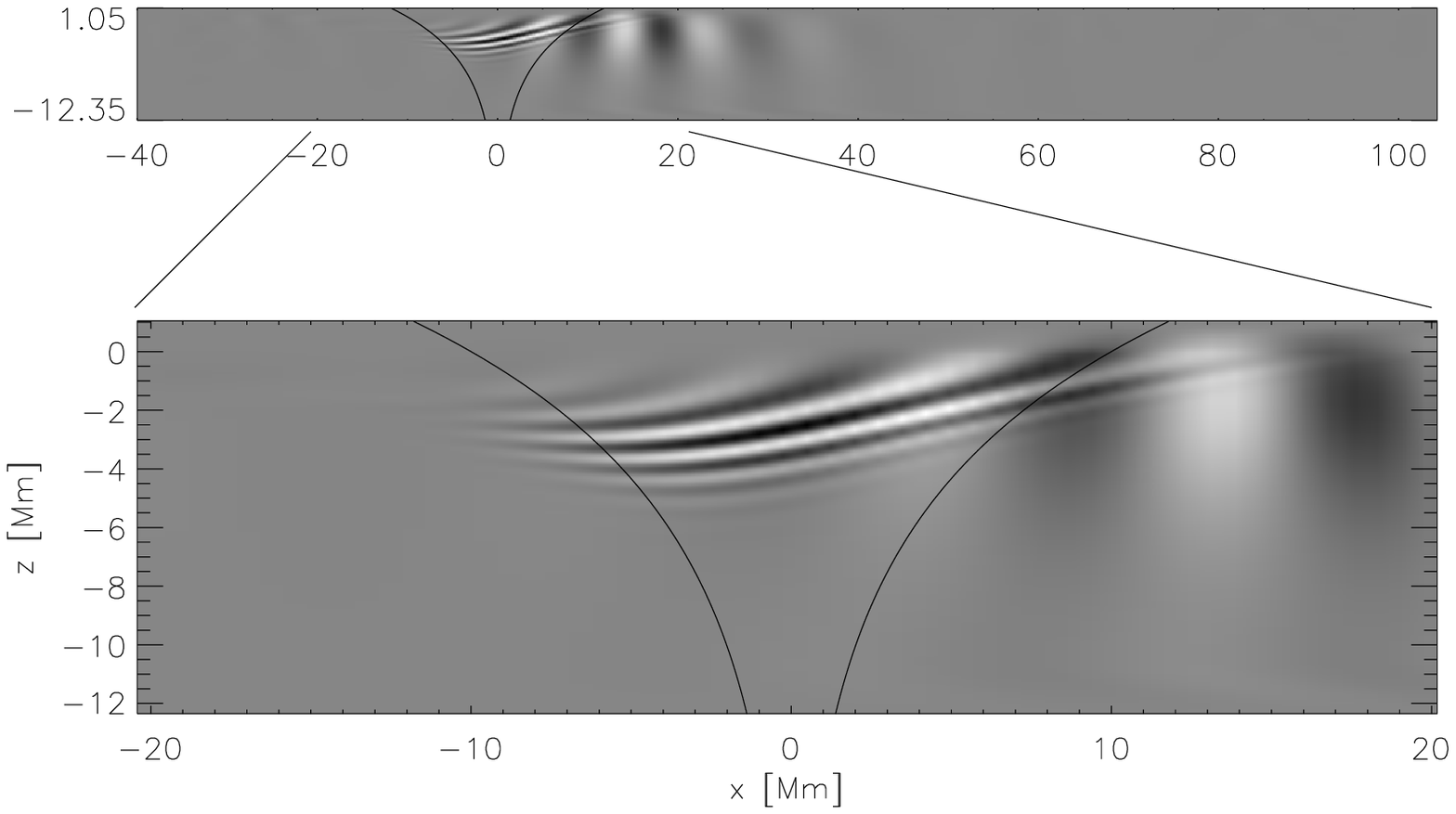,width=12.cm,clip=,angle=0.}
\caption{This plot shows $\rho_0 v_x'$ in the $x$--$z$ plane, through the sunspot axis.
The upper frames are for time $t=34$~min (before the {\it{f}}-mode  wave packet crosses the sunspot) and 
the lower frames are for $t=84$~min (after). 
The black curves with equation $B_z(r,z)=B_z(r=0,z)/2$ give an estimate of the ``width'' of the sunspot.
The conversion of the incoming {\it{f}} modes into slow magnetoacoustic modes is evident in the lower frames. 
}
\label{fig:modeconversion1}
\end{center}  
\end{figure}

First we consider a sunspot model with $\bs=3$~kG.  The simulation enables
us to understand what happens below the solar surface. The theory of the interaction of solar waves with 
sunspots had been developed using ray theory and two-dimensional numerical 
simulations (\eg ~\opencite{Cally07}; \opencite{Bogdan96}).  We find what appears to be very similar physics
in our three-dimensional simulations: the strong ``absorption'' of the {\it{f}} modes is consistent with 
partial conversion of the incoming waves into slow magnetoacoustic waves which
propagate along the field lines. This can be seen in Figure~\ref{fig:modeconversion1}.
The downward propagating slow magnetoacoustic modes 
experience a decrease in their speed as they propagate downwards and are therefore
shifted to increasingly short wavelengths.
Mode conversion is a very robust feature of this type of simulation.
We note that the slow magnetoacoustic waves are much easier to see in the $x$-component of 
wave velocity than in $v_z'$, which is why Figure~\ref{fig:modeconversion1} shows the former.

Since we are strongly damping short wavelengths, these magneto-acoustic modes rapidly decay. 
Any upward propagating 
wave encounters the damping buffer situated above the photosphere and is also damped.
In principle this decay is unphysical in both instances, however since neither the 
downward nor upward propagating waves return to the surface this is not undesirable.


\subsection{Wave Forms}
\label{sec.waveform}

In this section we compare the simulated $v_z'$ at $z=0$ with the observed cross-covariance.  
We emphasize that at the moment this comparison is not expected to be appropriate in
the immediate vicinity of the sunspot.
\begin{figure}    
\begin{center}
\psfig{file=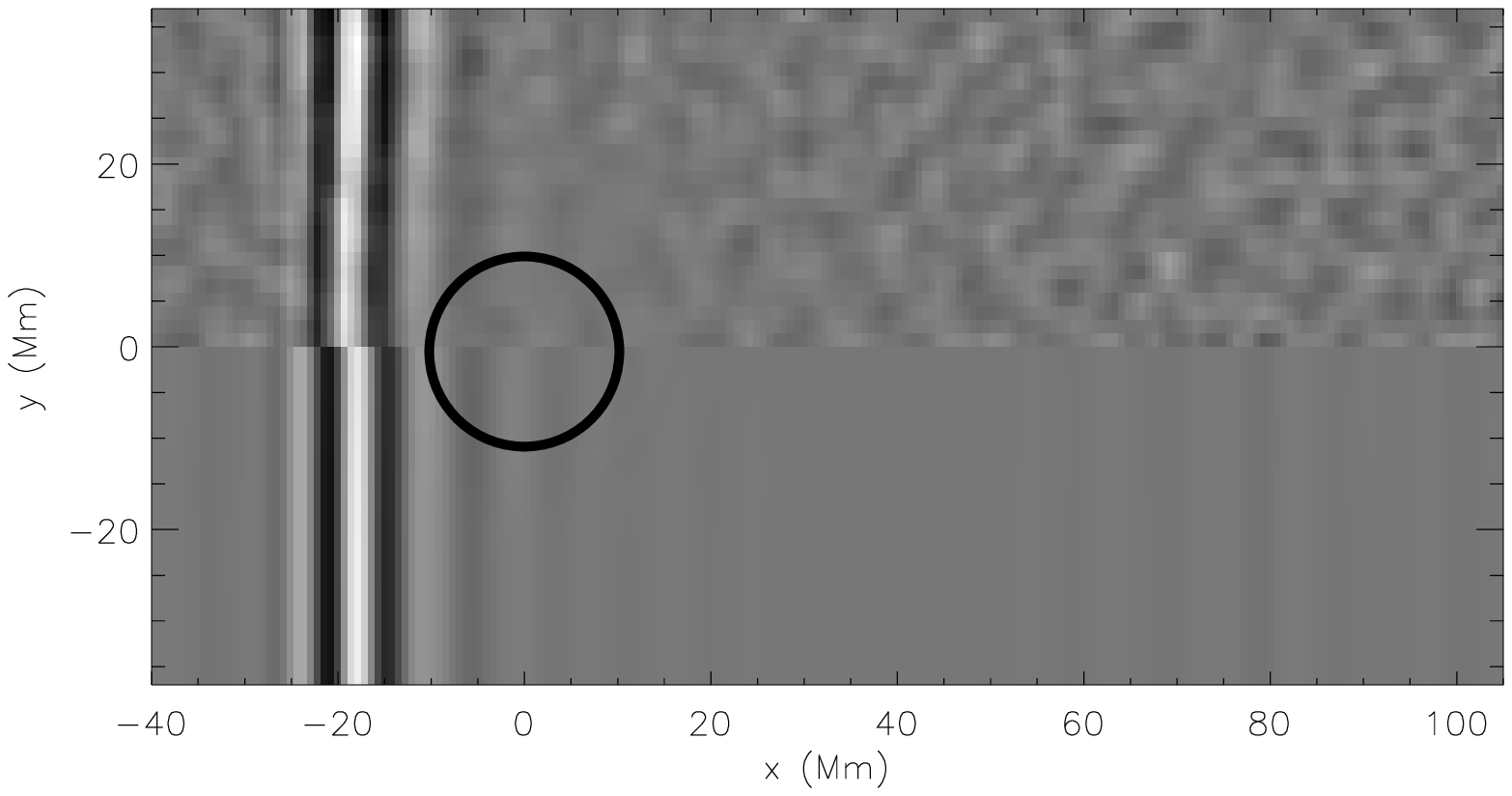,width=12.cm,clip=,bb=10 50 504 300,angle=0.}
\psfig{file=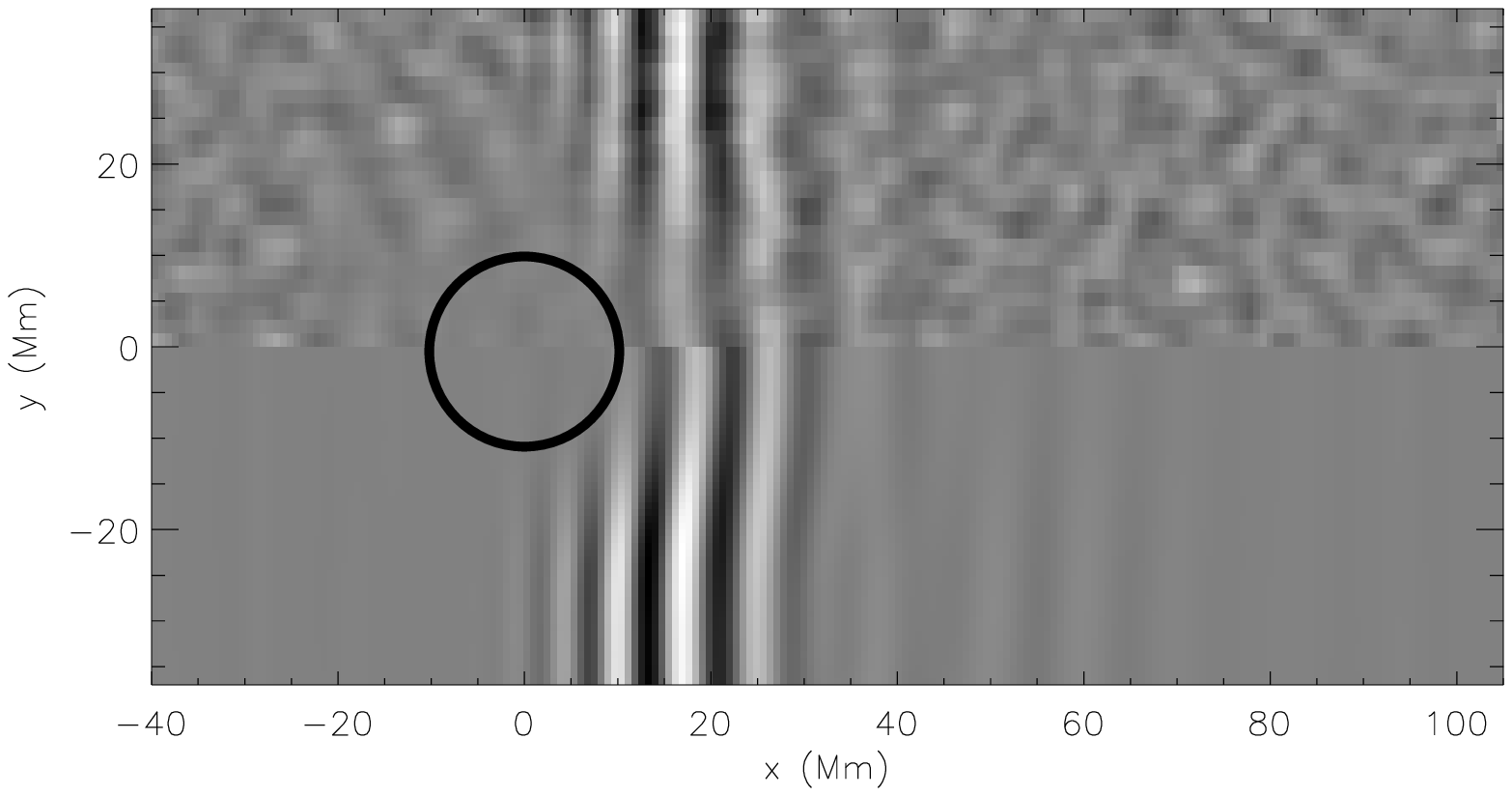,width=12.cm,clip=,bb=10 50 504 300,angle=0.}
\psfig{file=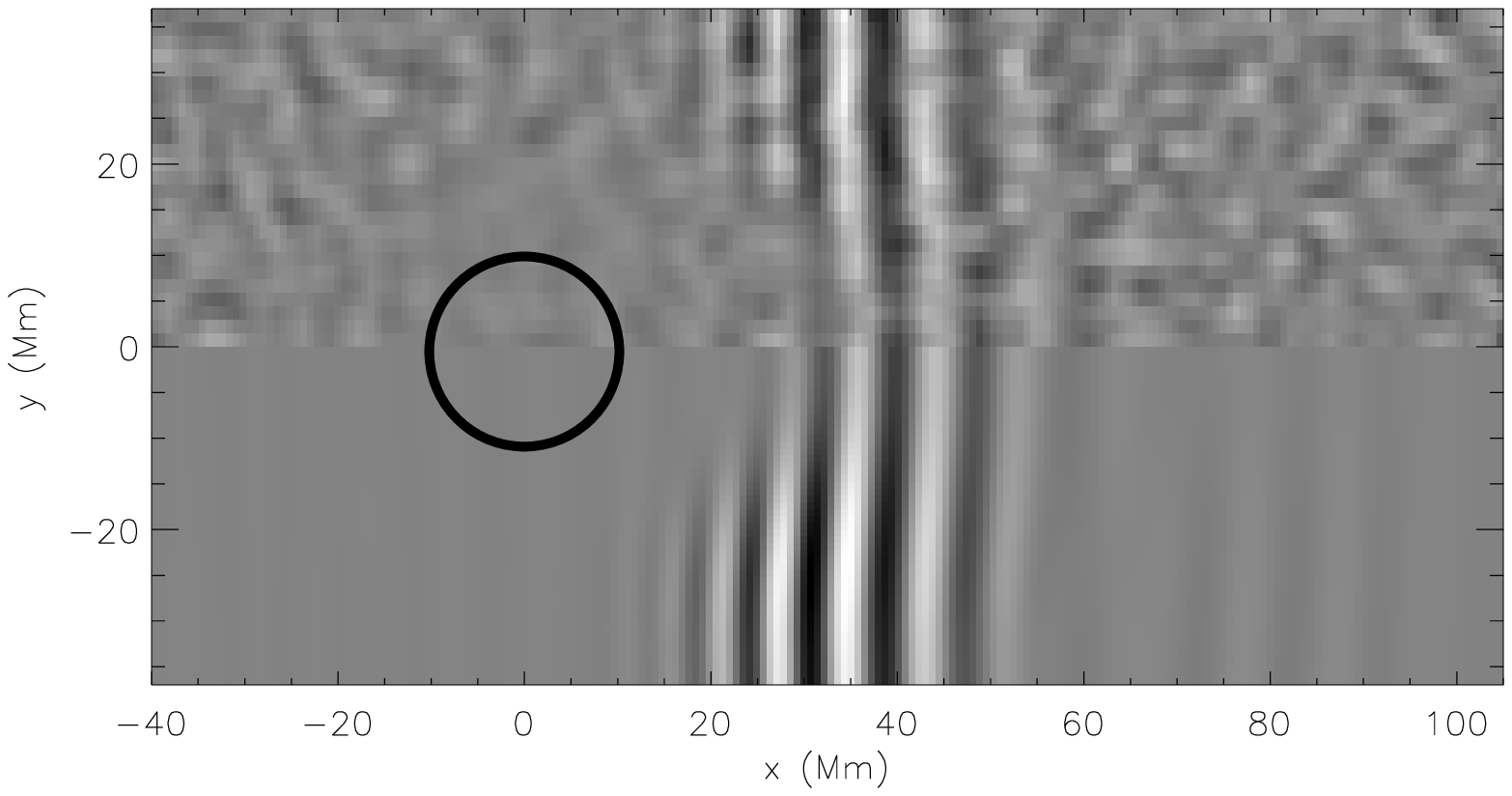,width=12.cm,clip=,bb=10 50 504 300,angle=0.}
\caption{Comparison of the simulated vertical velocity and the observed cross-covariance. In each panel
the upper frame shows the observed cross-covariance, the bottom panel the
simulated wave packet. The panels correspond to $t=40$~minutes, $t=100$~minutes, 
and $t=130$~minutes, from top to bottom. 
The black circles of radius $R_0=10$~Mm indicate the location of the sunspot.}
\label{fig:simvobs1-3}
\end{center}  
\end{figure}

Figure~\ref{fig:simvobs1-3} shows such a comparison between simulations and observations for times $t=40$~minutes,
$t=100$~minutes and $t=130$~minutes. The peak field strength used in this simulation 
was $\bs=3$~kG and the radius $R_0=10$~Mm. 
The comparison appears to be very good at time $t=130$~minutes, at which time the wavepacket has completely traversed 
the sunspot. At this time, all aspects of the waveform would appear to have been approximately 
reproduced by the simulation: amplitudes (including ``absorption''), phases, and spatial spectrum. 
The match appears to be less good for $t=40$~minutes; this will be discussed below.

\begin{figure}    
\begin{center}
\psfig{file=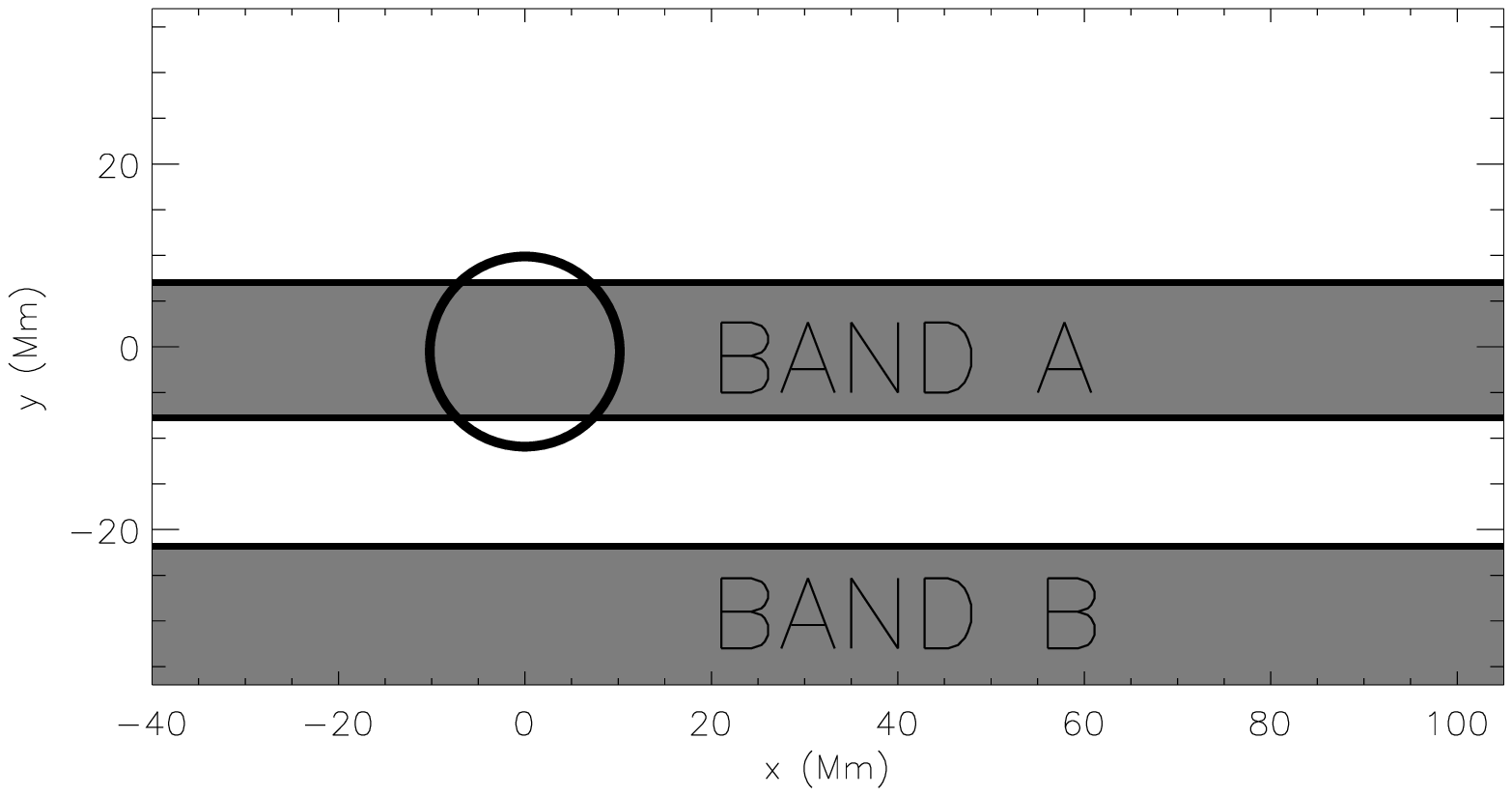,width=12.cm,clip=,angle=0.}
\caption{Sketch of the bands over which the data have been averaged in the $y$ direction in later plots. Band A
is centered on the spot, while band B acts as a reference.}
\label{fig.bandcartoon}
\end{center}  
\end{figure}

To be more precise, however,
the match between the simulations and observations must be better quantified. 
In order to reduce the observational noise, we have averaged the cross-covariance function 
in the $y$-direction over two bands.
Both bands, shown in Figure~\ref{fig.bandcartoon}, are $14.5$~Mm wide in the $y$-direction. 
The bands are labeled A and B and centered around $y=0$ and $y=\pm29$~Mm. Band A
is centered on the spot, while band B acts as a reference. 
The simulations are averaged in the same manner for comparison with the observations.

\begin{figure}    
\begin{center}
\psfig{file=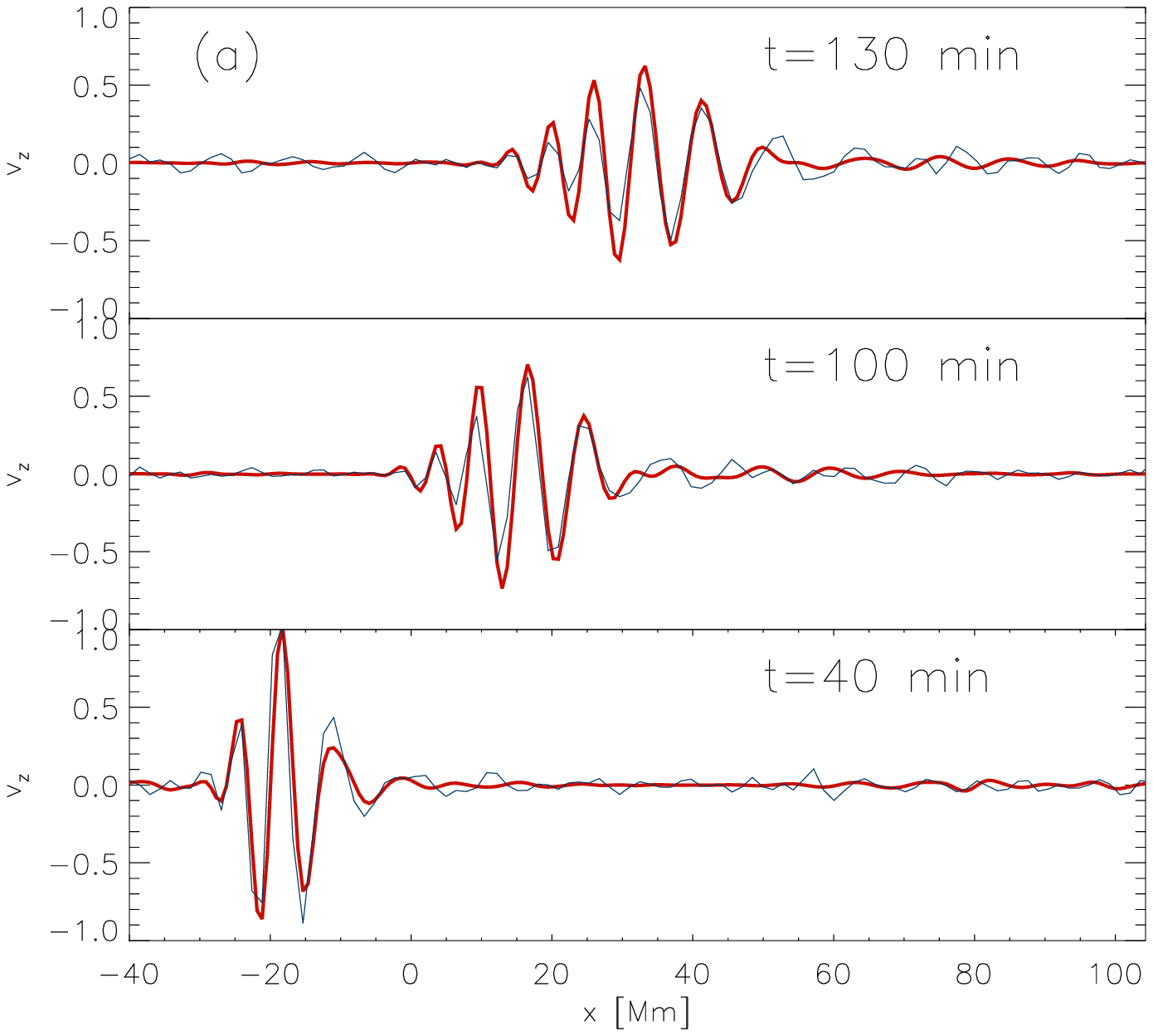,width=12.cm,clip=,angle=0.}
\psfig{file=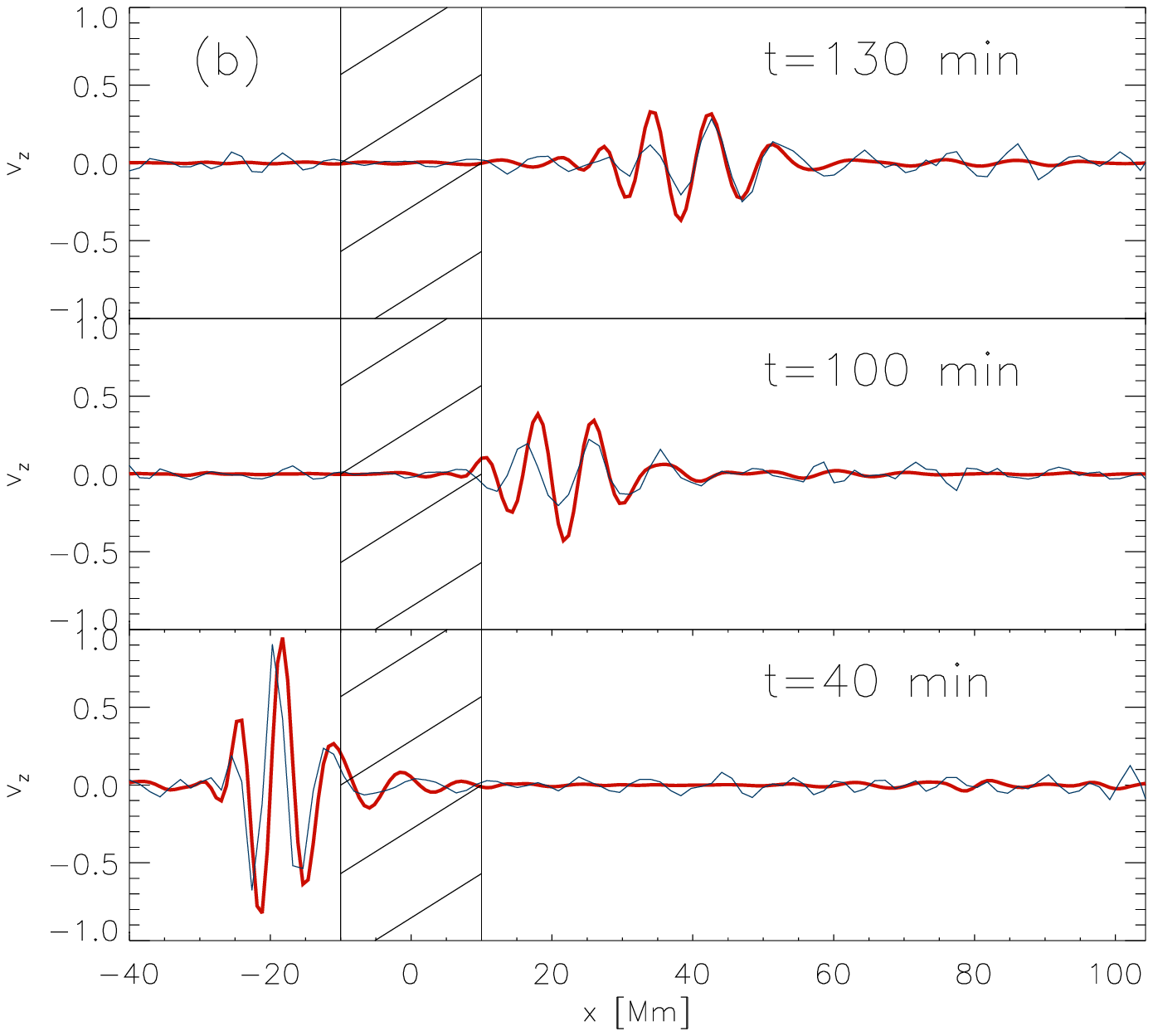,width=12.cm,clip=,angle=0.}
\caption{The top panels show the simulation ($v_z'$, thick red lines) and the observed 
cross-covariance ($C$, thin blue lines) averaged in the $y$-direction across band B (reference), 
at three different times ($t$). The bottom panels show the simulation ($v_z'$, $\bs=3$~kG, 
thick red lines) and observed cross-covariance ($C$, thin blue lines) averaged in the $y$-direction 
across band A, at three different times ($t$). The effect of the sunspot is very easily seen, 
by comparing with band B.}
\label{fig:bands}
\end{center}  
\end{figure}

In Figure~\ref{fig:bands} we have plotted both the simulation ($v_z'$, thick red
lines) and the observed cross-covariance ($C$, thin bluelines). 
In the top panels we compare the wave propagation at the edge of the 
computational box, \ie in band B. This part of the wave is little affected
by the sunspot  and the match between the simulation and observation is quite good since the 
initial amplitude spectrum $A_k$ was chosen accordingly.
In the  lower panels of  Figure~\ref{fig:bands} we see the results for the waves passing 
through the spot, \ie band A. 
First we notice that the wave amplitude is remarkably well reproduced in the simulations, meaning
that the observed wave ``absorption'' is consistent with mode conversion.
For $t=130$~minutes, after the waves have crossed the sunspot, there is no appreciable
phase shift between the simulation and the observations.
The match, however, is somewhat worse at $t=40$~minutes and there is an obvious
phase shift between the two wave forms with the observations lagging behind the simulations.
Given the value of the phase shift, we suspect that it is due to the effect of the 
moat flow. The moat flow is a horizontal outflow from the sunspot, which we have measured by tracking 
the small moving magnetic features. The observed moat velocity (averaged over nine days) has a peak value of $230$~m\,s$^{-1}$ 
at a distance of 25~Mm from the center of the sunspot and vanishes at a distance of $45$~Mm.
The solar waves moving through the flow are first slowed down (against the flow) and later sped up again (with the flow). 
We have not modeled this effect yet. 
It is reassuring, however, to see that the Doppler shift caused by the moat appears to have disappeared 
at $t=130$~minutes.

Reiterating, it appears that, by using numerical 
simulations, seismic signatures can be followed through their passage 
across the spot.

\section{Constraining $\bB$}
\label{sec.b}
\begin{figure}    
\begin{center}
\psfig{file=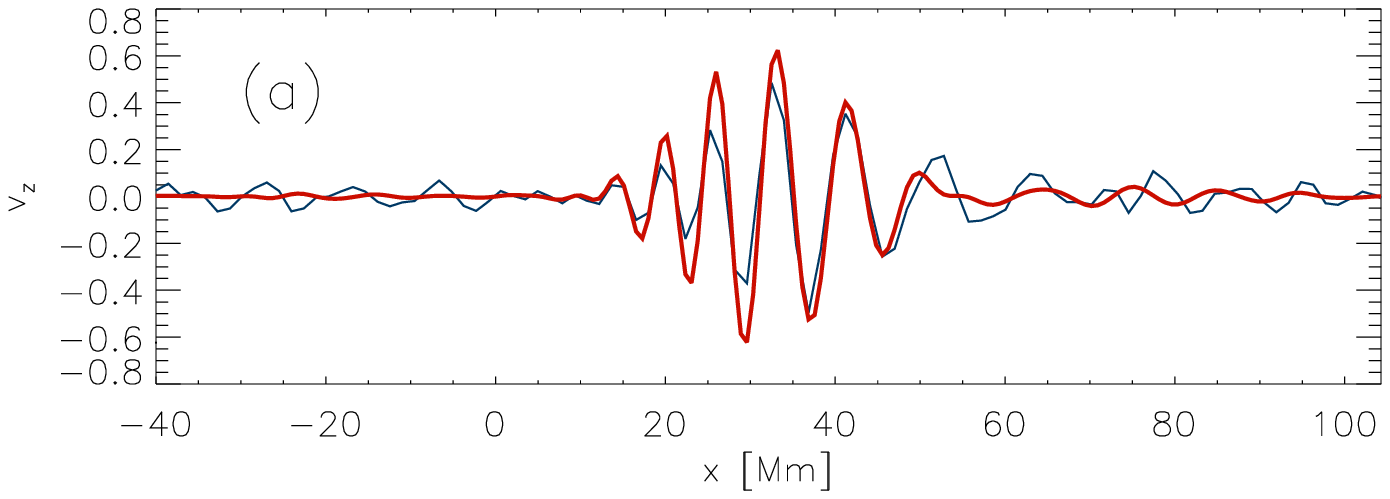,width=12.cm,clip=,bb=50 0 470 160,angle=0.}
\psfig{file=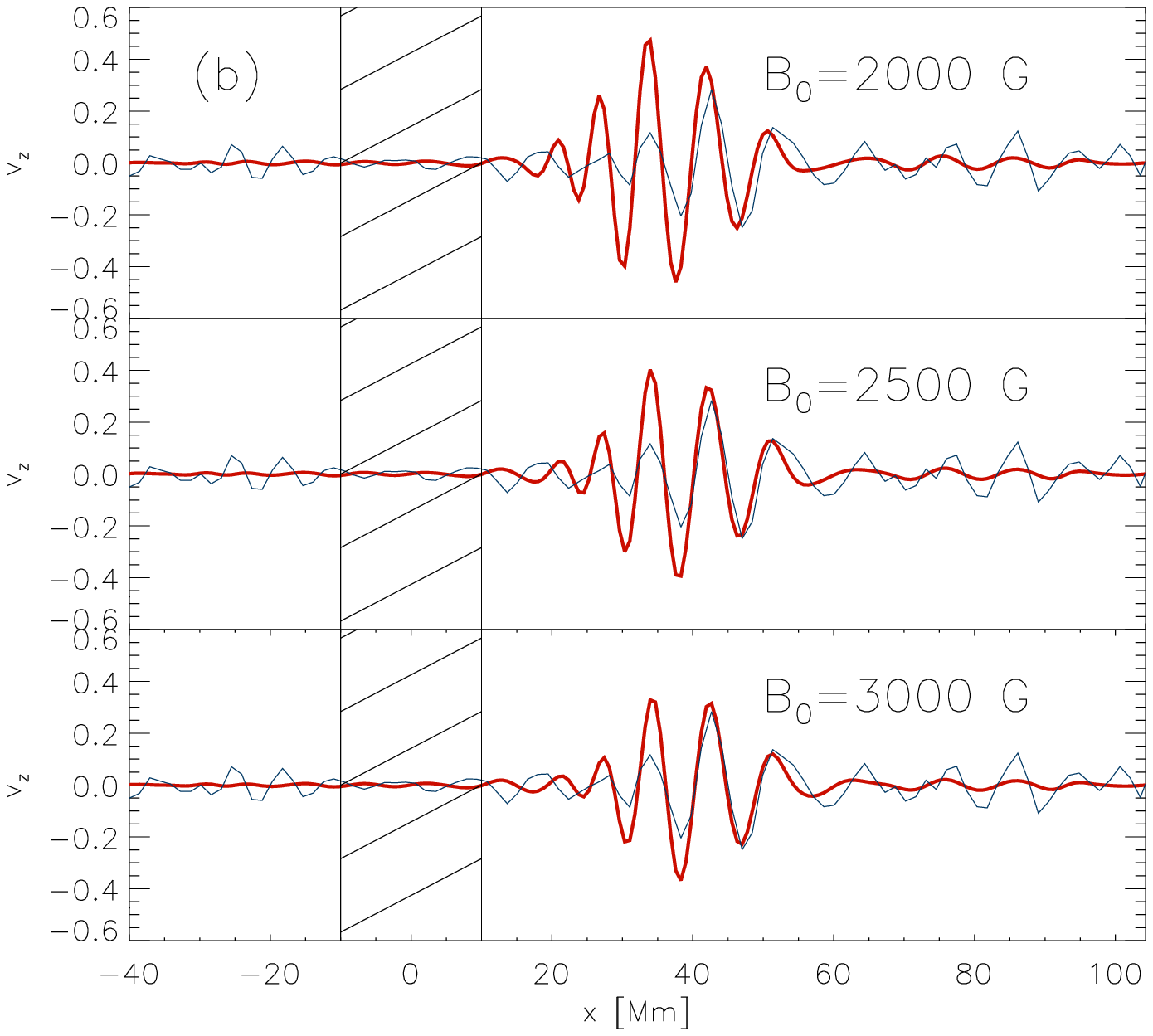,width=12.cm,clip=,bb=50 0 470 400,angle=0.}
\caption{Simulated vertical velocity (thick red lines) and observed cross-covariance (thin bluelines) at $t=130$~minutes.
(a) For the band at the edge of the domain (large $y$).
(b) For the band passing through the sunspot ($y=0$). The three panels correspond to simulations using $\bs=2$~kG, $2.5$~kG, and 
$3$~kG. It is seen that the simulations with $\bs=3$~kG provide the best match in 
terms of both amplitude and phase. The stripes indicate the location of the sunspot.}
\label{fig:b0_study_bands}
\end{center}  
\end{figure}

The solutions shown thus far have been for a model sunspot with a peak vertical
field strength $\bs=3$~kG at the surface $z=0$. The match is good, which raises
the question of whether other field strengths would match as well. The answer to
this question can be seen in Figures~\ref{fig:b0_study_bands}
 where we show the comparison between the 
simulations and observations for different field strengths, $\bs=2$~kG and $\bs=2.5$~kG. The comparison is
made some time after the wave packet has passed through the sunspot, so the
issue of the moat flow does not arise. At this stage we restrict ourselves to commenting that
qualitatively the match is best, in phase and amplitude, 
for $\bs=3$~kG;  for the other values of $\bs$ there
is an apparent phase mismatch along $y=0$ and the amplitude of that part
of the wave passing through the spot is not sufficiently damped. 
The quantification of the ``goodness
of fit'' between the simulation and observations, which will allow a more accurate 
determination of the field strength, will be the subject of a future study.

\section{Discussion}
\label{sec.discussion}
We have performed three-dimensional MHD simulations of waves propagating
through sunspot models. 
 The computations are set up in such a way as to allow comparing observed cross-covariances
(except in the immediate vicinity of the sunspot).
The parameters of the sunspot model can be chosen in such a way that its helioseismic signature is 
in good agreement with helioseismic observations of sunspot AR\,9787.

 A qualitative study using {\it{f}} modes has enabled us to place a constraint, $\bs\ge3$~kG, on the sunspot's 
near-surface field strength.  The remaining differences reflect real differences between the model 
and the observed sunspot, such as the moat flow. 

The model atmosphere that we have constructed also supports {\it{p}} modes with a dispersion relation that 
is very close to that of the Sun. Obviously, the {\it{p}} modes in combination with the {\it{f}} modes should enable 
us to place more substantial constraints on the subsurface structure of the sunspot. In particular, 
it should be possible to simultaneously constrain the magnetic field strength $\bs$, the sunspot 
radius $R_0$, and the magnetic  field inclination (controlled by the parameter $\alpha$). Of course, 
the parametric sunspot model that we have considered in this paper is just one particular model: we 
plan to consider other types of sunspot models in the future.

In summary, we believe that we have shown that the full-waveform modeling of sunspots is feasible.

\noindent

\acknowledgements 
SOHO is a project of international collaboration between ESA and NASA. We are grateful to Manfred Sch\"ussler 
for insightful discussions.

\end{article} 
\end{document}